\RequirePackage{fix-cm}
\documentclass[smallextended]{svjour3}       
\smartqed  
\usepackage{braket}
\usepackage{pgfplots}
\usepackage{csquotes}
\usepackage{float}
\usepackage{multirow}
\usepackage{dcolumn}
\usepackage[colorlinks=true, citecolor=blue]{hyperref}
\usepackage{hhline}
\usepackage{amssymb}
\begin{document}

\title{Quantum Robots Can Fly; Play Games: An IBM Quantum Experience
}


\author{Soumik Mahanti \and
        Santanu Das \and Bikash K. Behera \and Prasanta K. Panigrahi 
}


\institute{Soumik Mahanti \at
              Department of Physical Sciences, Indian Institute of Science Education and Research Kolkata, Mohanpur 741246, West Bengal, India\\
            \email{soumikmh1998@gmail.com}           
           \and
           Santanu Das \at
              Department of Mathematical Sciences, Indian Institute of Science Education and Research Kolkata, Mohanpur 741246, West Bengal, India\\
              \email{dassantanu315@gmail.com}
             \and
             Bikash K. Behera \at
              Department of Physical Sciences, Indian Institute of Science Education and Research Kolkata, Mohanpur 741246, West Bengal, India\\       \email{bkb13ms061@iiserkol.ac.in}   
              \and 
              Prasanta K. Panigrahi \at   Department of Physical Sciences, Indian Institute of Science Education and Research Kolkata, Mohanpur 741246, West Bengal, India \\ \email{pprasanta@iiserkol.ac.in} 
}

\date{Received: date / Accepted: date}

\maketitle

\begin{abstract}
Quantum Robot is an excellent future application that can be achieved with the help of a quantum computer. As a practical example, quantum controlled Braitenberg vehicles proposed by Raghuvanshi \emph{et al.} [Proceedings of the 37th International Symposium on Multiple-Valued Logic (2007)] is a mobile quantum system and hence acts as a quantum robot. Braitenberg vehicles are simple circuit robots which can experience natural behaviours like fear, aggression and love etc. These robots can be controlled by quantum circuits incorporating quantum principles such as entanglement and superposition. Complex behaviours can be mimicked by a quantum circuit that can be implemented in a quantum robot. Here we investigate the scheme of Raghuvanshi \emph{et al.} and propose a new quantum circuit to make the quantum robot fly. We demonstrate one of its application in playing a game. The quantum robot we present here shows the behaviour of `fear' and its movement is deterministic in nature. This phenomenon can be successfully modelled in a game, where it can always avoid accident. The proposed quantum circuit is designed in IBM quantum experience describing the above protocol.  
\end{abstract}

\keywords{Braitenberg Vehicles, Quantum Robot, Quantum Game, IBM Quantum Experience}

\section{Introduction}
\hspace{4mm} The compelling application of quantum-enabled technology predicts the dominance of quantum computers over the world in near future \cite{qrob_Nielsenbook}. Quantum computers can be used in the the area of robotics \cite{qrob_Craigbook2005} which is gradually becoming one of the major focuses of quantum computation and quantum information. A quantum robot is a mobile quantum system equipped with a quantum computer and ancilla system \cite{qrob_BenioffPRA1998}. Its major goal consists of specific changes in the state of the environment and to do measurement on it. Quantum robotics is currently drawing attention rapidly in the past few years. The main aim of quantum robotics is to apply quantum mechanics, quantum computation and quantum algorithms in the study of robotics. On the other hand, the application of robotics and engineering principles can be used to develop quantum computers and implement it as well \cite{qrob_Quantumroboticsgroup}. A beautiful yet strange domain of quantum information and quantum computation is stretching its wings in the premise of commercialization. However, it is not far away from now that robots controlled by quantum processors are going to come in handy in research and commercial world on a regular basis. Those kind of robots will be able to use the resource of entanglement, superposition principle of quantum mechanics \cite{qrob_RaghuvanshiIEEE2007}, violation of Bell's inequality \cite{qrob_RoweNature2001} and will open a new horizon of robotics technology and their ability \cite{qrob_RaghuvanshiIEEE2007}. 
 
The first idea of theoretical quantum robot was introduced by Benioff \cite{qrob_BenioffPRA1998,qrob_BenioffarXiv0003006} on a strictly quantum world. Then Raghuvanshi \emph{et al.} \cite{qrob_RaghuvanshiIEEE2007} showed some practical examples of quantum controlled robots which used the idea of Braitenberg vehicles \cite{qrob_Braitenbergbook1986}. They used classical sensors and classical motors with a quantum circuit (or it can be thought like a quantum brain) that decides the movement of the quantum vehicle \cite{qrob_Raghuvanshirp2006}. For a binary input with only two possible situations, the quantum circuit used there would determine the movement of the robot probabilistically and had three degrees of freedom in general such as moving to right, left and straight; while in this present work, we add one more degree of freedom to the robot as it can fly and would deterministically select its path. We use light with particular intensity and two different input states correspond to whether light is falling on the sensors or not.

We design a quantum circuit and simulate it by using IBM Quantum Experience. Recently, IBM has developed different types of prototypes of quantum processors and they are available by a free web based interface called \textit{IBM Quantum Experience} (IBM QE). Researchers have taken a proper advantage of it by demonstrating and running a variety of quantum computing experiments, e.g., \cite{qrob_AggarwalarXiv:1804.08655v12018,qrob_SrinivasanarXiv:1805.109282018,qrob_DasharXiv1710.051962017,qrob_VishnuarXiv:1709.05697,qrob_SatyajitarXiv:1712.05485,qrob_RoyarXiv:1710.10717,qrob_GangopadhyayQIP2017,qrob_BeheraarXiv:1712.008542017,qrob_JhaarXiv:1806.10221,qrob_DasharXiv:1805.10478,qrob_BeherarXiv:1803.06530,qrob_SrinivasanarXiv:1801.00778,qrob_GurnaniarXiv:1712.10231,qrob_HarperarXiv:1806.023592018,qrob_HegadearXiv:1712.073262017,qrob_KapilarXiv:1807.00521,qrob_MohantaarXiv:1807.00323,qrob_BeherarXiv:1806.10229,qrob_MS1,qrob_Wooton}. Here, we use IBM's 5-qubit quantum processor, `ibmqx4' to design the quantum circuit and perform the experiment by simulating it. We verify the new protocol of quantum robot and illlustrate its application in a game.   
 
The paper is organized as follows. Sec. \ref{qrob_Sec2} discusses the theoretical scheme of a Braitenberg vehicle in brief. Secs. \ref{qrob_Sec3} \& \ref{qrob_Sec4} focuses on the quantum controlled circuit and how it simulates the behaviour of the robot deterministically. Sec. \ref{qrob_Sec5} explicates an application of quantum robot in a game, where our robot is a perfect winner. Finally, we conclude the paper in Sec. \ref{qrob_Sec6} by providing its future implications and applications.        

\section{The Classical Braitenberg Vehicle \label{qrob_Sec2}}
  
\begin{figure}[]
\includegraphics[scale=1]{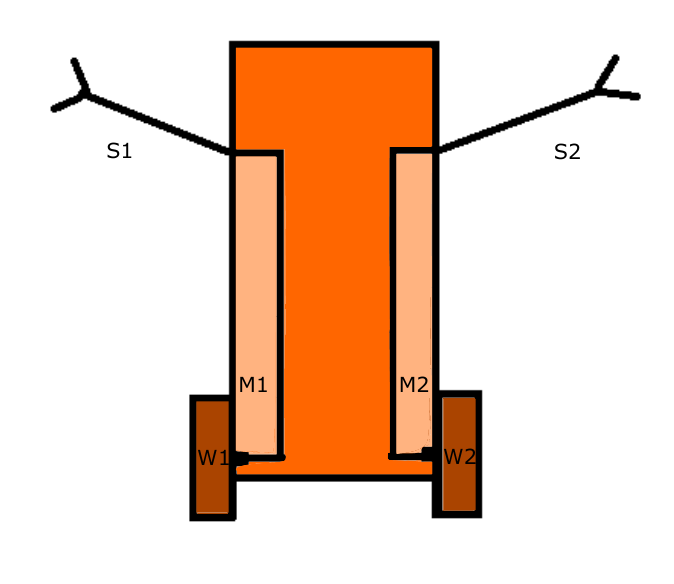}
\caption{\textbf{Schematic Side View of Braitenberg Vehicle}. The light sensors S1 and S2 are presented here as an angle which are connected to the motors M1 and M2 respectively, represented by rectangular boxes. Here W1 and W2 are wheels of the vehicle.}
\label{qrob_Fig1}
\end{figure}

\hspace{4mm} The revolutionary and striking idea behind Braitenberg vehicles \cite{qrob_Braitenbergbook1986} is that with only simple circuit structure some mixed and complex behaviour can be shown. A schematic diagram of a simple Braitenberg vehicle is depicted in Fig. \ref{qrob_Fig1}. 

\begin{figure}[]
\includegraphics[scale=1]{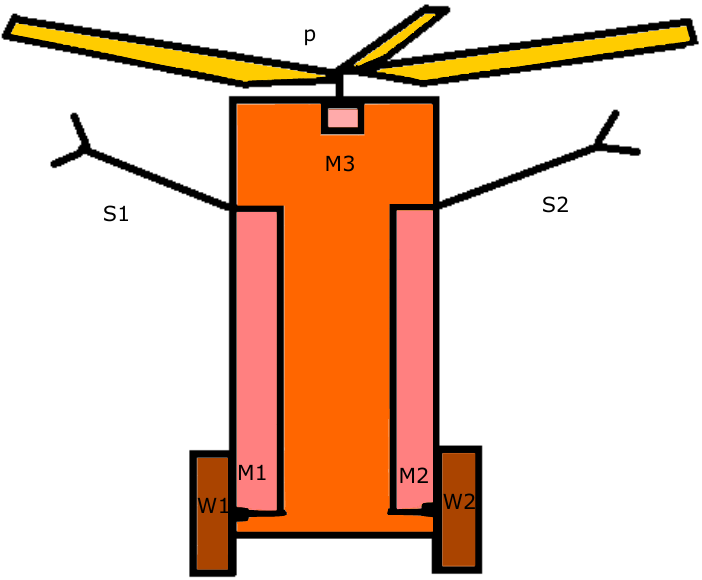}
\caption{\textbf{Schematic Side View of Quantum Robot}. The light sensors S1 and S2 are presented here as an angle which are connected to the motors M1 and M2 respectively and the motor M3 is connected with a propeller P.}
\label{qrob_Fig2}
\end{figure}

As observed from Fig. \ref{qrob_Fig1}, the two light sensors, S1 and S2 are directly connected with the two motors, M1 and M2 respectively. The mechanism is such that if a light sensor senses some light, then the motor connected to it starts and in absence of light, the motor turns off. When the robot is exposed to high light intensity, both the motor is turned on and as a result, the robot moves forward at full speed. In the case when only the left light sensor, S1 senses some light, only the left motor, M1 starts and the motor moves towards right hand side. The same goes for the case of light coming from right hand side in which case the motor turns to the left. This behaviour of the robot can be interpreted as `fear' or `shyness'. In our work, we have added one more degree of freedom to the simple Braitenberg vehicle. We present a vehicle that has two light sensors (S1 \& S2) attached with it and two motors (M1 \& M2) which are connected to the sensors as depicted in Fig. \ref{qrob_Fig2}. However, here we have used another additional motor, M3 which makes the robot fly in certain situation. The design of the quantum circuit is such that it deterministically controls the behaviour of these kind of vehicles. The schematic diagram of our quantum controlled Braitenberg vehicle is illustrated in Fig. \ref{qrob_Fig2}. 

\section{The Quantum Scheme \label{qrob_Sec3}} 
\hspace{4mm} The robot designed here has three motors, M1, M2 and M3 and two sensors, S1 and S2. We are denoting $\Ket{0}$ as light is not falling on sensors, and $\Ket{1}$ as light is falling on the sensors. The two motors (M1 \& M2) of robot are connected directly to the two sensors (S1 \& S2) as shown in Fig. \ref{qrob_Fig2}. We introduce a third motor, M3 that is attached with a propeller, P overhead of the vehicle. When the third motor (M3) is active, the vehicle takes off from the ground. The rest of its actions are same as the classical model discussed earlier in Sec. \ref{qrob_Sec2}. All the motors (M1, M2 \& M3) used in this robot have binary output as $\Ket{0}$ and $\Ket{1}$. Here $\Ket{0}$ denotes that the motor is not moving and $\Ket{1}$ denotes that the motor is moving. We design a quantum circuit such that when the input of both the sensors (S1 \& S2) are $\Ket{00}$, the output of motors (M1\& M2) is $\Ket{11}$ that implies the robot moves forward. When the input of both the sensors (S1 \& S2) is $\Ket{01}$ then the output is $\Ket{01}$ that means the right motor (M2) runs and the left motor (M1) remains off, so that the robot can take a turn towards the left and the robot turns away from the source. When the input of both the sensors (S1 \& S2) is $\Ket{10}$ then the output is $\Ket{10}$ and similarly, it avoids closeness to the source. When the input of both the sensors (S1 \& S2) is $\Ket{11}$, the output is $\Ket{00}$ and in this situation the third motor starts and the robot takes a flight. The details of the movement of the robot is tabulated in Table \ref{qrob_Tab1}. 

\begin{table}[]
\centering
\begin{tabular}{ c c c c c c }
\hline
\hline
$S1^{\dagger}$ & $S2^{\ddagger}$ & M1$^{||}$ & M2$^{\perp}$ & M3$^{\star}$ & Quantum Robot's Behavior\\
\hline
0 & 0 & 1 & 1 & 0 &  Moves forward\\
0 & 1 & 0 & 1 & 0 &  Takes a left turn\\
1 & 0 & 1 & 0 & 0 &  Takes a right turn\\ 
1 & 1 & 0 & 0 & 1 &  Takes off from the ground\\
\hline
\hline
\end{tabular}\\
$\dagger$ First sensor, $\ddagger$ Second sensor, $||$ First motor, $\perp$ Second motor, $^{\star}$ Third motor connected to propeller . 
\caption{\textbf{The table details the robot behaviour upon the sensation of light.}}
\label{qrob_Tab1}
\end{table} 

\section{The Quantum Circuit \label{qrob_Sec4}}

\subsection{Working and Decomposition of Two-Control-Two-Target Toffoli ($C^{2}(X_{3}X_{4})$) Gate} 
\begin{figure}[H]
\includegraphics[scale=.7]{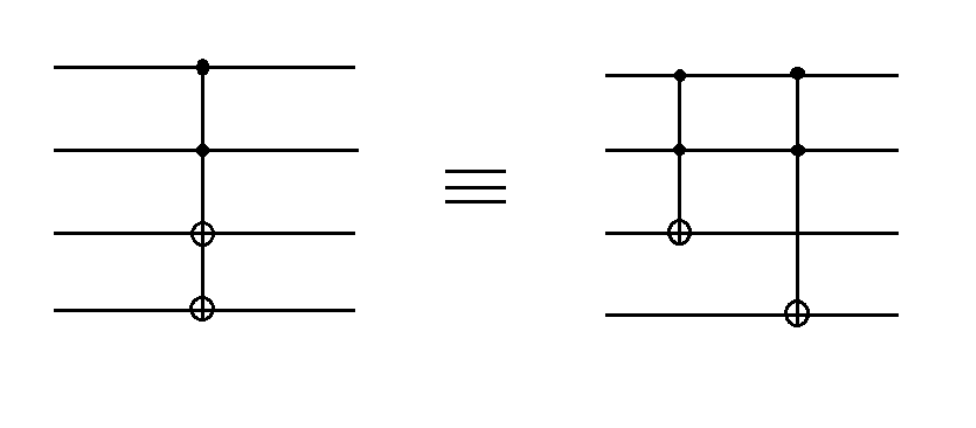} 
\caption{\textbf{Equivalent quantum circuit for two-control-two-target Toffoli ($C^{2}(X_{1}X_{2}$) gate.}}
\label{qrob_Fig3}
\end{figure}

\hspace{4mm} The working of two-control-two-target Toffoli gate $C^{2}(X_{3}X_{4})$ is described as follows. There are two control qubits and two NOT operations in $C^{2}(X_{3}X_{4})$ gate. A four-qubit quantum state is taken as an input. The gate flips both the third and fourth qubit of the state only in case when the first two qubits are in $\ket{11}$ state, otherwise there is no change in the the output state. The decomposition of $C^{2}(X_{3}X_{4})$ is illustrated in Fig. \ref{qrob_Fig3}. Two Toffoli gates ($C^{2}(X_{3})$ \& $C^{2}(X_{4})$) are used to design the above quantum gate. 

\subsection{Circuit diagram}
\begin{figure}[]
\includegraphics[scale=.7]{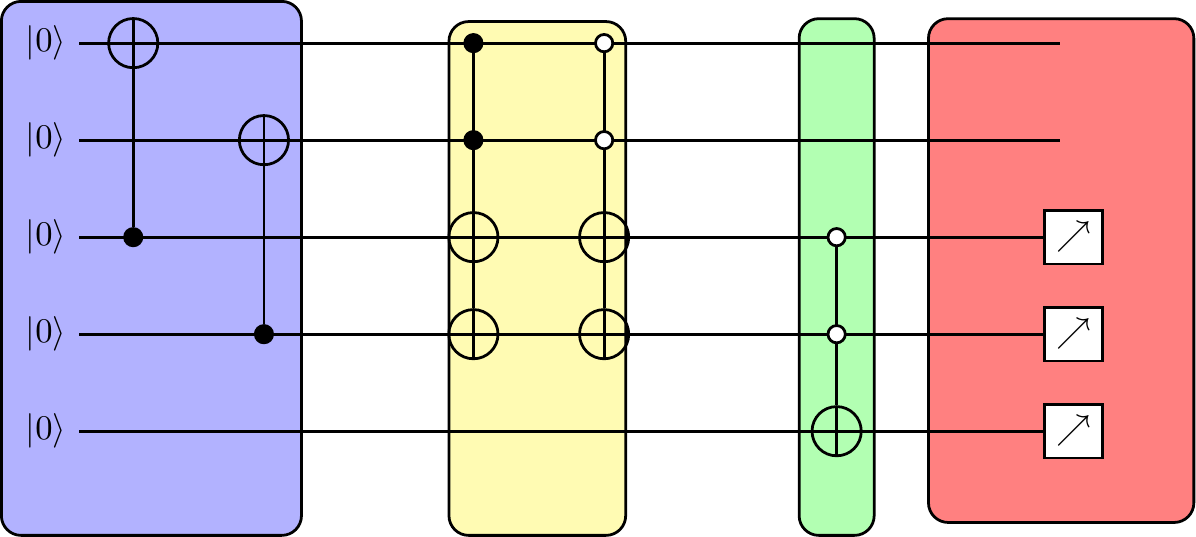} 
\caption{\textbf{Quantum circuit for implementing the proposed protocol of the quantum robot.}
The top two qubits represent two ancilla qubits, the next two are the input qubits and the last one is attached with the third motor (M3). Measurements are done in the Z basis on the last 3 qubits, which determines the activation of the certain motors and thus controls the movement of the robot.}
\label{qrob_Fig4}
\end{figure}

\hspace{4mm} Here we use IBM Quantum Experience's 5-qubit quantum chip, `ibmqx4' for performing the experiment. The first two qubits, q[0] and q[1] (See Figs. \ref{qrob_Fig4} \& \ref{qrob_Fig5}) represent the ancillary qubits, and the last three qubits, q[2], q[3] and q[4] correspond to three motors M1, M2 and M3 respectively. 

\begin{itemize}
\item If the input is $\Ket{00}$, then the initial state of the whole system is $\Ket{00000}$. After the action of two CNOT gates on two ancilla qubits (q[0] \& q[1]), the output becomes $\Ket{00000}$. Then the two-control-two-target Toffoli gate with the controlled qubits (q[0] \& q[1]) as $\Ket{11}$ followed by a two-anticontrol-two-target Toffoli gate with controlled qubits (q[0] \& q[1]) $\Ket{00}$, act on the input qubits (q[2] \& q[3]) and the state of the whole system becomes $\Ket{00110}$. Finally, a Toffoli gate with two anticontrol qubits (q[2] \& q[3]) operates on the three qubits (q[2], q[3] \& q[4]) connected to the motors (M1, M2 \& M3) and hence the system's state is changed to $\Ket{00110}$. Then measurement along the Z-basis on the last three qubits (q[2], q[3] \& q[4]) gives the result as 1, 1 and 0 for motors M1, M2 and M3 respectively. Thus, both the motors (M1 \& M2) connected with wheels (W1 \& W2) start and the quantum robot moves forward, while the third motor (M3) remains inactive. 

\item Similarly for the input $\Ket{01}$, the initial state of the whole system is $\Ket{00010}$. In a similar manner by action of gates, it can be seen that the final state of the system remains $\Ket{01010}$. Then measurement along the Z-basis on the last three qubits (q[2], q[3] \& q[4]) gives the result as 0, 1 and 0 for motors M1, M2 and M3 respectively. Hence only the motor connected with the right wheel (W2) starts and the quantum robot takes a left turn. 

\item With the input state $\Ket{10}$, then the initial state of the whole system is $\Ket{00100}$ and the final state becomes $\Ket{10100}$. Then measurement along the Z-basis on the last three qubits (q[2], q[3] \& q[4]) gives the result as 1, 0 and 0 for motors M1, M2 and M3 respectively. Hence only the motor (M1) connected with the left wheel (W1) starts and the quantum robot turns right.

\item The interesting case happens for the input $\Ket{11}$. Then the initial state of the whole system is $\Ket{00110}$ and the system's state is finally changed into $\Ket{11001}$. Then measurement along the Z-basis on the last three qubits (q[2], q[3] \& q[4]) gives the result as 0, 0 and 1 for motors M1, M2 and M3 respectively. Here both the motors (M1 \& M2) connected with wheels (W1 \& W2) remains turned off but the third motor (M3) becomes active now, and the robot takes off from the ground.
\end{itemize}

\subsection{Simulation of the Circuit Diagram}
\begin{figure}[H]
\includegraphics[scale=.45]{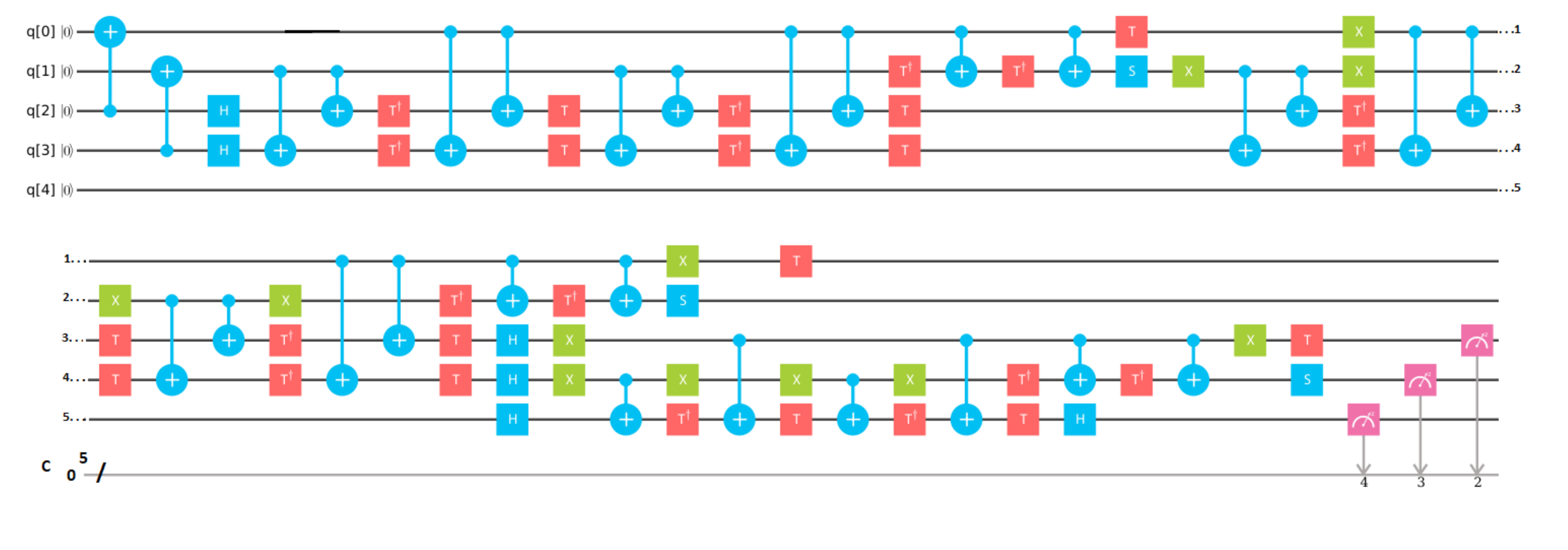}
\centering
\caption{\textbf{Construction of the quantum circuit in IBM QE}. Two-controlled two-target Toffoli gate (shown in yellow colored box of Fig. \ref{qrob_Fig4}) is decomposed into parallel c Fonnection of two two-controlled Toffoli gates using 1,2,3 and 1,2,4 qubit respectively. The two controlled Toffoli gates are decomposed into Hadamard, S, T, $T^{\dagger}$ gates as explained in Ref. \cite{qrob_Nielsenbook}.}
\label{qrob_Fig5}
\end{figure}

We constructed our quantum circuit in IBM Quantum Experience using the quantum gates such as CNOT, Hadamard (H), X, T, $T^{\dagger}$, S, $S^{\dagger}$. Since two-controlled-two-target Toffoli gates do not have a standard notation in IBM-Q Experience, they are constructed by using standard gates such as Hadamard, X, T, $T^{\dagger}$, S, $S^{\dagger}$. The result of our simulation is presented in Fig. \ref{qrob_Fig5}.
\begin{figure}[H]
\includegraphics[scale=.71]{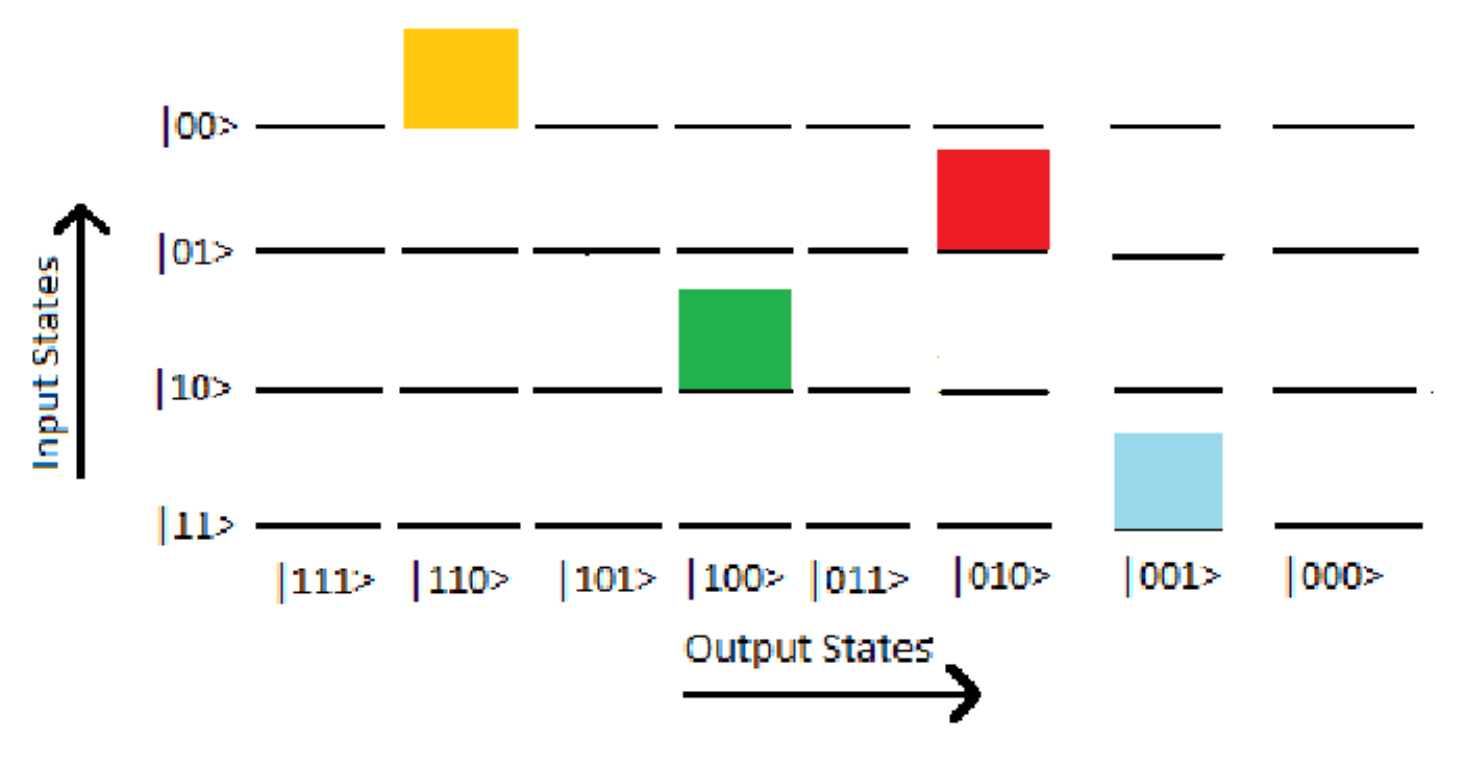}
\caption{\textbf{Result of Simulation of the Quantum Circuit}.
For a particular input, the output state is deterministic and is represented by the colored boxes whose size determines the probability of that particular output state. In our case, the size of each box is unity. For example, the yellow colored box for input state $\Ket{00}$ represents that only the output state of $\Ket{110}$ appears with probability one and other output states does not show up at all.}
\label{qrob_Fig6}
\end{figure}

\section{Application of Quantum Robot in a Game \label{qrob_Sec5}}
\hspace{4mm} The model of quantum robot introduced here has an advantage over a simple game. Its schematic diagram is presented in Fig. \ref{qrob_Fig6}. Following are the two assumptions to play the game. 

\begin{figure}[H]
\includegraphics[scale=1.3]{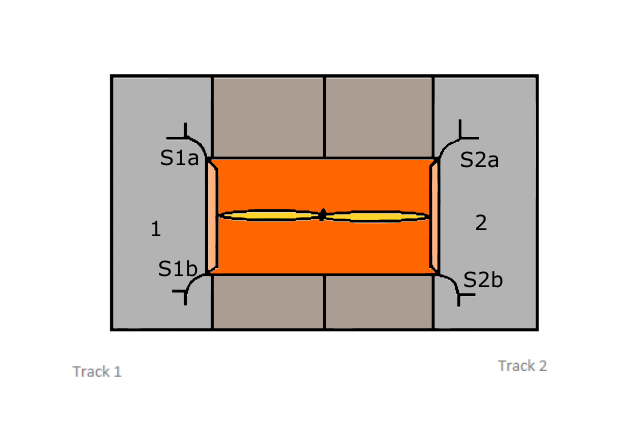}
\caption{\textbf{Advantage of Quantum Robot Over A Game.} Top view of the quantum robot in a hypothetical game situation is depicted. Here 1 and 2 are the tracks allowed for obstacles. S1a and S1b are sensors on one side whose input goes through an OR-gate to the first input of the quantum circuit. S2a and S2b are similarly connected to the second input.}
\label{qrob_Fig7}
\end{figure}

\begin{itemize}
\item This is a four lane road where there is no divider in between them. The robot spreads over two lanes and it is always in motion with a constant speed. The motors M1 and M2 of the robot are designed such that they help it to change lane upon receiving proper signals. 

\item There are various other obstacles (which sends some kind of light signals) moving on only the two terminal tracks named as Track 1 and Track 2 (See Fig. \ref{qrob_Fig7}). They move randomly on both ways up or down, but all of them have same uniform speed with respect to the robot. There is one starting point and one finishing line.
\end{itemize}

The rule of the game is described as follows. The quantum robot has to cover the distance upto the finishing line without touching any of the obstacles. Here we use the same kind of robot but it would have four sensors (S1a, S1b, S2a \& S2b) and each couple of sensors on one side would be connected through an OR gate to the input of that side of the vehicle. If from any side there is some obstacle, the input to the circuit will be $\Ket{1}$. If the right motor turns on, the robot will shift a lane towards the left and vice-versa. While both the motors are turned on, the robot remains still. When obstacles are coming from both the tracks, then the third motor (M3) connected to propeller helps the robot to take off from the ground. Now in this hypothetical construction of a game, our robot will always win it. The condition of winning is apparent from the previous Table \ref{qrob_Tab1} that represents the motion of the robot in different scenarios.
 
\section{Conclusion \label{qrob_Sec6}}
\hspace{4mm} To conclude we have demonstrated here a quantum algorithm for a quantum Braitenberg vehicle, where one more motor has been added that enables the vehicle to fly. A new quantum circuit has been proposed to manipulate the behaviour of the quantum robot and has been simulated using the IBM Q Experience platform. Furthermore, the concept of quantum robot proposed here has been applied to a simple game to avoid accident in the game. Advancement in the area of robotics is one of the major concerns of scientists since the very beginning of technological development \cite{qrob_Craigbook2005}. When it comes to quantum robot, its importance grows much more because of various surprising and strange new phenomena operating in the quantum domain. Benioff's \cite{qrob_BenioffPRA1998} idea of quantum robot has been applied in our physical world to make a quantum robot and now the challenge is to extend its ability and versatility. Our goal of this project is to cultivate the ideas further to have a quantum emotional robot in near future. The construction of robot with quantum brain explains well how a quantum system is interacting with the surroundings and other quantum phenomena \cite{qrob_BenioffPRA1998}. In future quantum technology, a quantum computer can be used for various purposes like robot action planning, problem solving and vision \cite{qrob_RaghuvanshiIEEE2007} to perform significantly faster than any modern robotics technology.         

\section*{Acknowledgments}
\label{acknowledgments}
\hspace{4mm} S.D. and S.M. acknowledge the support of INSPIRE fellowship, awarded by the Department of Science and Technology, Government of India. B.K.B. acknowledge the support of Institute fellowship provided by IISER Kolkata. S.D. and S.M. acknowledge Indian Institute of Science Education and Research Kolkata for providing hospitality during which a part of this work was completed. We are extremely grateful to IBM quantum experience project. The discussions and opinions developed in this paper are only those of the authors and do not reflect the opinions of IBM or any of it's employees.

\end{document}